\documentclass[referee]{raa}            

\usepackage{graphicx,times}             
\usepackage{natbib}
\usepackage{amssymb,amsmath}
\usepackage{epstopdf}
\bibpunct{(}{)}{;}{a}{}{,}

\usepackage[a4paper=true,pagebackref=true]{hyperref}
\hypersetup{colorlinks = true, linkcolor = green, anchorcolor = red, citecolor = blue, filecolor = red, pagecolor = red, urlcolor = red}

\def\fbs{\textsc{fbs}}

\begin{document}

\title{Long-period eclipsing binaries: towards the true mass-luminosity relation. %
I. The test sample, observations and data analysis.
}

   \volnopage{Vol.0 (20xx) No.0, 000--000}      
   \setcounter{page}{1}          

   \author{Alexei Yu. Kniazev
      \inst{1,2,3}
   \and Oleg Yu. Malkov
      \inst{4}
   \and Ivan Yu. Katkov
      \inst{3,5,6}
   \and Leonid N. Berdnikov
      \inst{3}
   }

   \institute{South African Astronomical Observatory, P.O. Box 9, Observatory, Cape Town, 7935 South Africa {\it aknizev@saao.ac.za}\\
         \and
	     Southern African Large Telescope, P.O. Box 9, Observatory, Cape Town, 7935 South Africa\\
	     \and
	     Sternberg Astronomical Institute, Lomonosov Moscow State University, Moscow, 119992 Russia\\
	     \and
	     Institute of Astronomy of RAS, Moscow, Russia\\
	     \and
          New York University Abu Dhabi, PO Box 129188, Abu Dhabi, UAE\\
         \and
         Center for Astro, Particle, and Planetary Physics, NYU Abu Dhabi, PO Box 129188, Abu Dhabi, UAE\\
\vs\no
   {\small Received~~20xx month day; accepted~~20xx~~month day}}

\abstract{%
The mass-luminosity relation is a fundamental law of astrophysics.
We have suggested that the currently used
mass-luminosity relation is not correct for the $M/M_\odot > 2.7$ range of mass
since it was created using double-lined eclipsing binaries, where the
components are synchronized and consequently change each other's evolutionary path. To exclude this effect we have started a project to study
long-period massive eclipsing binaries in order to construct radial velocity
curves and determine masses for the components.
We outline our project and present the selected test sample
together with the first HRS/SALT spectral observations
and the software package,
\fbs\ (Fitting Binary Stars), that we developed for the analysis of our spectral data.
{As the first result we present the radial velocity curves and best-fit orbital elements
for the two components of the FP\,Car binary system from our test sample.}
}

   \authorrunning{A. Yu. Kniazev, O. Yu. Malkov, I. Yu. Katkov, \& L. Berdnikov }  
   \titlerunning{Study of long-period eclipsing binaries}  

   \maketitle

\section{Introduction}           
\label{sect:intro}

The mass of a star is the parameter that, to a first approximation, is most important in
determining its evolution. However, the mass cannot be dynamically determined
for a single star, so indirect methods have been developed for estimating stellar masses.
The most widely used of them is to estimate the mass from observations of the distribution
of another parameter for a stellar ensemble under study (field stars, cluster stars). 
The stellar luminosity is the most commonly used parameter, and the subsequent transition
to masses of stars is made using the so-called mass-luminosity relation (MLR).

Independent determination of the mass of a star and its luminosity is only possible
for components of binary systems of certain types.

One suitable type of binary system is a
{visual binary star with known orbital parameters and trigonometric parallax.}
Such  stars are usually wide pairs, whose components do not interact with each other
and are evolutionarily similar to single stars. In addition, usually they are
in the nearest solar neighborhood and, therefore, are mostly low-mass stars.
The problem of determining the masses of visual binaries was discussed, for example,
in \citet{2016MNRAS.459.1580D,2012A&A...546A..69M,1998A&A...338..455F}.
The construction of the MLR for low-mass stars based on observational data is discussed
in \citet{2004ASPC..318..159H,2000A&A...364..217D,1999ApJ...512..864H,1997A&A...320...79M}.

Another major source of independently defined stellar masses is detached eclipsing binary stars
with components on the main sequence, where the spectral lines of both components
are observed (hereafter double-lined eclipsing binaries, DLEB). These stars are usually
relatively massive ($M/M_\odot > 1.5$) and their parameters are used to construct the
stellar MLR for intermediate and large masses. The exact parameters of DLEB stars
and the MLR based on them can be found, for example, in \citet{2010A&ARv..18...67T,2001ARep...45..972K,
1998ARep...42..793G,1991A&ARv...3...91A,1980ARA&A..18..115P}.

When these two MLRs (based on the visual binaries and on the DLEB stars with components
on the main sequence) are jointly analyzed and used (in particular, in order to compare
the theoretical MLRs with empirical data), it is generally assumed by default,
that the components of the detached close binaries and the wide binaries evolve in a similar way.
It should be noted, however, that DLEB are close pairs whose components' rotation
is synchronized by tidal interaction, and, due to rotational deceleration,
they evolve differently than ``isolated'' (i.e., single or wide binary systems) stars.

When comparing the radii of DLEB and single stars \citep{2003A&A...402.1055M},
a noticeable difference between the observed parameters of B0V--G0V components
of DLEB and of single stars of similar spectral classes was found. This difference
was confirmed by analysis of independent studies published by other authors.
This difference also explains the disagreement between the published scales
of bolometric corrections. Larger radii and higher temperatures of A-F components
of DLEB stars can be explained by the synchronization and associated slowing down of rotation
of such components in close systems. Another possible reason is the effect
of observational selection: due to the non-sphericity of the rotating stars,
the parameters determined from the observations depend on the relative orientations
of their rotation axes. Isolated stars are oriented randomly, while components
of eclipsing binaries are usually observed from near the equatorial plane.
Systematically smaller observed radii of DLEB stars of spectral class B can be explained
by the fact that stars with large radii do not occur with companions on
 the main sequence: most of them have already filled their Roche lobe
(which stopped their further growth) and have become semi-detached systems
(which excluded them from the discussed statistics).
Then, in~\citet{2007MNRAS.382.1073M} data for the fundamental parameters of the components
of a few currently known long-period DLEB have been collected. These stars
presumably have not undergone synchronization of rotation with the orbital period
and therefore spin rapidly, and evolve similarly to single stars.

{The theory of synchronization (and circularization) in close
binary systems developed by \citet{1975A&A....41..329Z, 1977A&A....57..383Z}
is based on the mechanism of energy dissipation
via dynamic tides in non-adiabatic surface layers of the component
stars. Another theory was developed by \citet{1987ApJ...322..856T, 1988ApJ...324L..71T},
and is based on tidal dissipation of the kinetic energy of large-scale meridional flows.
In their critical reviews, \citet{2007MNRAS.382..356K, 2010MNRAS.401..257K} point out
that both the circularization and synchronization time-scales implied by these mechanisms
differ by almost three orders of magnitude,
and,  based on an analysis of the observed rates of apsidal motion, show that
the observed synchronization times agree with Zahn's theory but are inconsistent with the
shorter time-scale proposed by Tassoul.

The synchronization time depends
primarily on the stellar mass and the binary separation.
So, for example, according to \citet{1987ApJ...322..856T},
the synchronization time for orbital periods up to about 25 days
is smaller than one-tenth of the main-sequence life-time of a 
$3 \: M_\odot$ star.

The theories of synchronization mentioned above have been developed
for early-type, massive stars with radiative envelopes (i.e., for stars with
$M/M_\odot > 1.5$).
In the current work, to construct the mass-luminosity relation for ``isolated'' stars,
we study DLEB stars in the range $M/M_\odot > 2.7$,
as the masses of components of other types of binary stars (visual binaries,
resolved spectroscopic binaries) rarely exceed this limit
(stars in the range $1.5 < M/M_\odot < 2.7$ will be considered later).
According to the shorter time-scale theory of Tassoul,
the synchronization time becomes comparable to the main-sequence life-time
of a $2.7 \: M_\odot$ star for orbital periods of the order of 50-70 days
(the longer time-scale theory of Zahn predicts even shorter periods).}

Currently there is no way to properly estimate the degree
to which the effect on the IMF may be important for $M > 2.7 \: M_\odot$,
as available observational data for that mass range are too poor to
draw definite conclusions.
For that reason we have started a pilot project to study
 long-period massive eclipsing binaries to construct radial velocity
curves and determine the masses of their components.
Using published photometric data or light-curve solutions
we expect to obtain luminosities for the individual components.
With accurate luminosity determinations
we plan to compare their location on the mass-luminosity diagram
with the ``standard'' MLR.
As a result of this pilot study we plan to confirm that
rapid and slow rotators satisfy different MLRs,
which should be used for different purposes.
Then the feasibility
of a larger project, the construction of a reliable
``fast rotators'' MLR, will be considered. The data we obtain will also
be used to establish mass-radius and mass-temperature relations.

\section{The Test Sample, Observations and Data reduction}
\label{sect:Obs}

\begin{table}[thb]
\begin{center}
\caption[]{The Test Sample.}\label{Tab:Sample}
\begin{tabular}{ccllrcr}  \hline\hline
\# &   Name   & RA (2000.0) &DEC (2000.0)&   mag &  e   &Period   \\
(1)&   (2)    &    (3)      &   (4)      &  (5)  & (6)  & (7)     \\
\hline
01 & V883 Ara & 16:51:45.10 &-50:17:46.5 &  8.55 & $\gg$0 & 61.8740 \\
02 &   KV CMa & 06:50:52.67 &-20:54:37.4 &  7.16 & $\gg$0 & 68.3842 \\
03 & V338 Car & 11:13:52.31 &-58:36:30.4 &  9.30 & $=$0 & 74.6429 \\
04 & V884 Mon & 07:05:11.84 &-11:06:02.4 &  9.13 & $=$0 &123.2100 \\
05 & V766 Sgr & 17:51:57.00 &-28:17:02.0 & 10.80 & $=$0 &147.1050 \\
06 &   FP Car & 11:04:35.87 &-62:34:22.2 &  9.70 & $=$0 &176.0270 \\
07 &V1108 Sgr & 19:12:43.63 &-18:08:12.0 & 11.50 & $\gg$0 & 46.5816 \\
08 &   PW Pup & 07:49:06.00 &-31:07:42.6 &  9.20 & $=$0 &158.0000 \\
09 &   mu Sgr & 18:13:45.81 &-21:03:31.8 &  3.80 & $\gg$0 &180.5500 \\
10 &   AL Vel & 08:31:11.28 &-47:39:57.4 &  8.60 & $=$0 & 96.1070 \\
11 &   NN Del & 20:46:49.22 &+07:33:10.4 &  8.39 & $\gg$0 & 99.2684 \\
\hline\hline
\end{tabular}
\end{center}
\end{table}

To compile the test sample for our pilot project we used
the Catalog of Eclipsing Variables
\cite[hereafter CEV;][]{2007A&A...465..549M,2013AN....334..860A,2014MNRAS.444.1982A} from which
we have carefully selected 11 massive long-period
(i.e., presumably non-synchronised) detached main-sequence
eclipsing systems that are presented in Table~\ref{Tab:Sample}.
The selected systems should have components of similar luminosities
(i.e., can be observed as SB2 systems -- spectroscopic binaries, 
where spectral lines from both components are visible)
and guarantee an accurate determination of stellar parameters
(in particular masses to 3\%) of early-type stars composing them.
We planned to obtain a minimum of five spectra for targets with circular orbits ($e=0$)
and a minimum of 10 spectra for targets with non-circular orbits ($e \gg 0$).
This number of spectra should be enough to find a
credible orbital solution and to complete the science objectives.

All observations were obtained with 
the High Resolution Spectrograph \citep[HRS;][]{Ba08, Br10, Br12, Cr14}
at the Southern African Large Telescope \citep[SALT;][]{Buck06,Dono06}.
The HRS was used in the medium resolution (MR) mode, that gives a spectral resolution
R$\sim$36\,500--39\,000; it has an input fiber diameter of 2.23 arcsec for both object and sky.
All our \'echelle data were obtained during 2017--2019 and cover the total
spectral range $\approx$3900--8900~\AA, where both blue and red CCDs were used with 1$\times$1 binning.
All science observations were supported by the HRS Calibration Plan,
which includes a set of bias frames at the beginning of each observational night
and a set of flat-fields and a spectrum of a ThAr lamp once per week.
Since the HRS is a vacuum \'echelle spectrograph installed inside a temperature-controlled
enclosure such a set of calibrations is enough to give an average external velocity accuracy of 300 m~s$^{-1}$ \citep{KUKB19}.
HRS data underwent a primary reduction with the SALT science pipeline \citep{Cra2010}, which includes
overscan correction, bias subtractions and gain correction.
After that \'echelle spectroscopic reduction was carried out using
the HRS pipeline described in detail in \citet{KGB16, KUKB19}.

\section{Fitting Binary Stars: full pixel fitting method}
\label{sect:analys}

To analyze fully reduced HRS spectra of binary systems and to determine stellar atmosphere parameters
for each component such as effective temperature $T_\mathrm{eff}$, surface gravity $\log g$,
metallicity [Z/H] as well as stellar rotation $v \sin i$ and line-of-sight velocities $V_j$
we developed a dedicated \textsc{Python}-based package,
Fitting Binary Stars (\fbs) (Katkov et al., 2020 in preparation).
The \fbs\ implements a full pixel fitting approach to simultaneous approximation of multiple
epoch spectra of binary system by combination of two synthetic stellar models
using $\chi^2$ minimization.
The \fbs\ developed on top of the non-linear minimization \textsc{lmfit}
package \citep{lmfit} provides a high-level interface to many optimization
methods (e.g. Levenberg-Marquardt, Powell, Downhill simplex Nelder-Mead method,
Differential evolution etc.).

During the evaluation of $\chi^2$ the \fbs\ proceeds by the following steps:
First, the \fbs\ interpolates two stellar templates from the grid of synthetic stellar spectra
for given sets of stellar atmosphere parameters (T$_\mathrm{eff}$, $\log g$, [Z/H])$_{1,2}$.
The interpolation has to be fast to work with high-resolution stellar spectra containing tens
to hundreds thousands of pixels in the spectral range under analysis.
Therefore, we propose an algorithm where the \fbs\ pre-calculates the Delaunay triangulation
in 3 dimensions of stellar model parameters ($T_\mathrm{eff}$, $\log g$, [Z/H]) using nodes
of the synthetic grid.
Interpolating the \fbs\ finds the simplex containing the given point, then averages the spectra
from the simplex vertices with weights inversely proportional to the squared distance to the vertex.
Such an algorithm is very fast and might work on regular as well as irregular model grids with missing nodes. 
Then, the model templates are broadened by individual stellar rotation {$v \sin i$}$_{1,2}$
and shifted for the line-of-sight velocities $V_1^j$, $V_2^j$ at the epoch of the $j$-th spectrum.
The last two steps are to sum templates with weights $w_{1,2}$ and multiply the spectrum
by the extinction curve appropriate to ithe assumed $E(B-V)$ or to multiply the final spectrum by a polynomial continuum
to match the difference between the observed and synthetic spectra.
In such an approach the $\chi^2$ value can be written as follows:
\begin{equation}
\chi^2 = \sum_j \chi^2_j = \sum_j \sum_\lambda \left( \frac{F^j_\lambda - M^j_\lambda}{\delta F^j_\lambda} \right)^2\\
\end{equation}

\begin{equation}
    M^j_\lambda = C_\lambda \sum_{k={1,2}} w_k \cdot S\left(T_{\mathrm{eff} k}, \log g_k, [Z/H]_k\right) * 
    \mathcal{L}(V^j_k, v \sin i_k),
\end{equation}
where $F^j_\lambda$, $\delta F^j_\lambda$, $M^j_\lambda$ represent the observed spectrum at the $j$-th epoch,
its uncertainties and the model, respectively; $S$ is the stellar template interpolated from the grid of stellar models;
$\mathcal{L}$ is the convolution  kernel to derive the broadening effect
due to stellar rotation \citep{Gray92} and to shift the templates by the line-of-sight velocity
of each binary star component $k$ at epoch $j$; ``$*$'' denotes convolution;
$C_\lambda$  is a polynomial multiplicative continuum or extinction curve for the given $E(B-V)$.
The approach produces the following parameters:
$T_\mathrm{eff}$, $\log g$, [Z/H], $v \sin i$ for both components of the binary star,
$n$ pairs of line-of-sight radial velocities $V^j_1$, $V^j_2$ for $n$ spectra observed
at different epochs and  $E(B-V)$ for the system.

Hereafter, we usualy employ \cite{Coelho14} stellar models
(stars with T$_{\mathrm{eff}} = 3000-26000$~K),
which match the HRS MR instrumental resolution \citep{KUKB19} well.
Also we adapted high-resolution \textsc{tlusty} models 
\citep[T$_{\mathrm{eff}} = 15000-55000$~K;][]{2003ApJS..146..417L,2007ApJS..169...83L}
and \textsc{phoenix} models \citep[T$_{\mathrm{eff}} = 3000-23000$~K;][]{phoenix},
convolving them to match the HRS MR instrumental resolution.

\begin{figure}
 \begin{minipage}[h]{0.490\linewidth}
   \centering
   \includegraphics[angle=0, width=\textwidth, clip=]{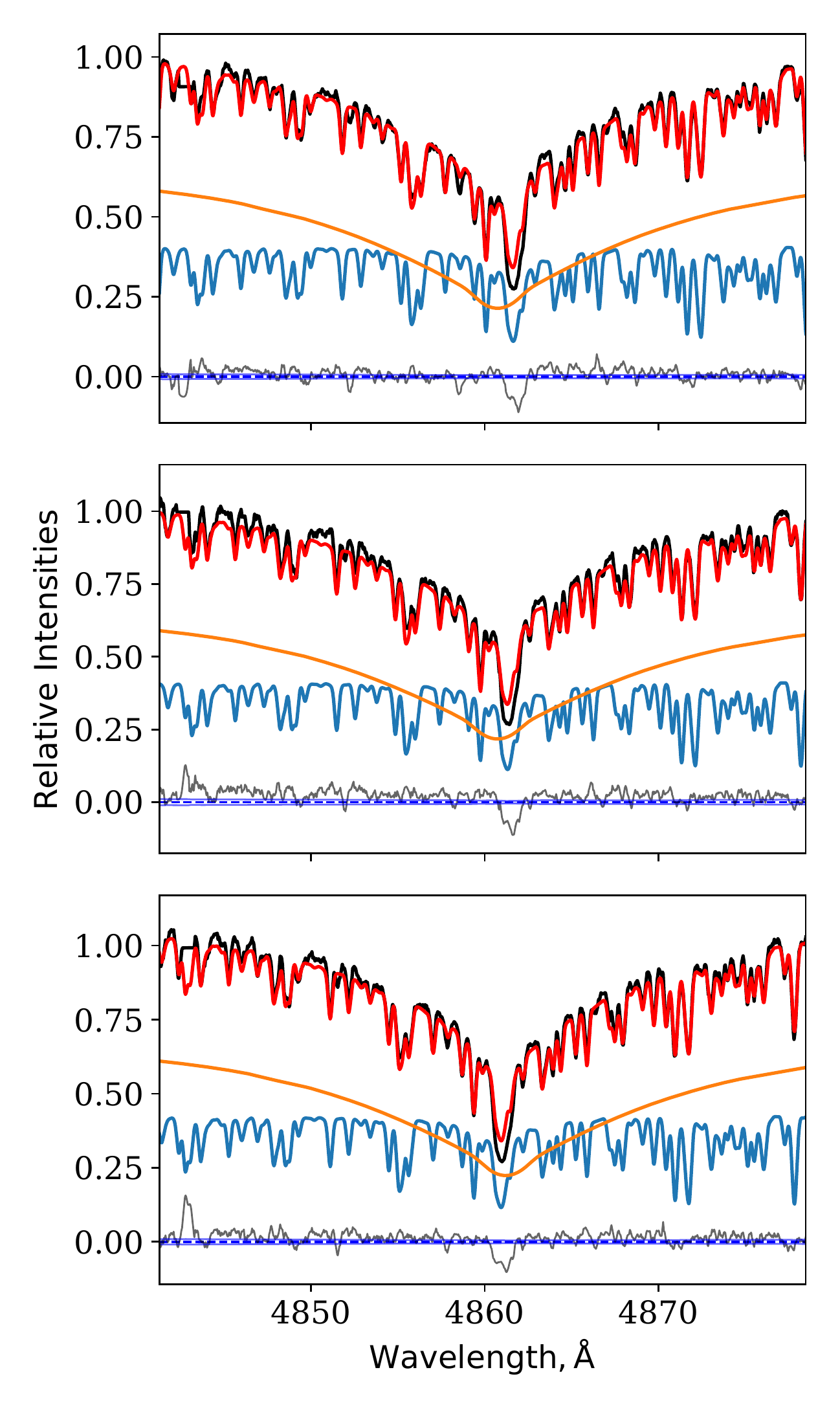}
   \caption{Example of the processing of three observed spectra of FP\,Car (three epochs) with the \fbs\ software.
   Each panel shows the part of the observed spectrum in the region of the H$\beta$ line in black.
   The result of modelling is shown in red.
   The two components are shown in blue and orange, respectively.
   The difference between the observed and modelled spectra is shown at the bottom of the panel in grey, together
	 with errors that were propagated from the HRS data reduction (continuous dark blue lines).\label{fig:examples}
   }
 \end{minipage}
%
 \begin{minipage}[h]{0.490\linewidth}
  \centering
   \includegraphics[angle=-90, width=\textwidth, clip=]{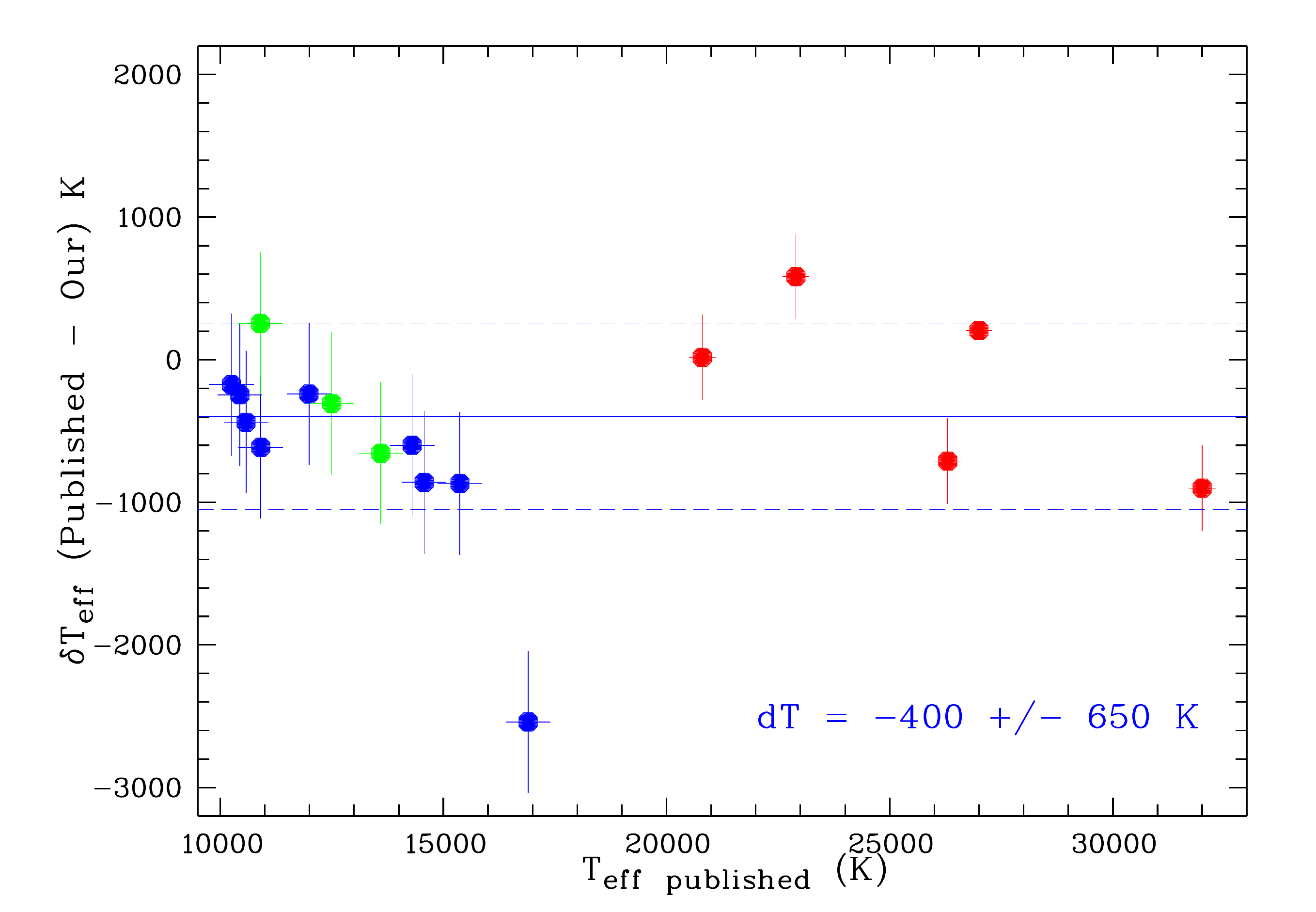}
   \includegraphics[angle=-90, width=\textwidth, clip=]{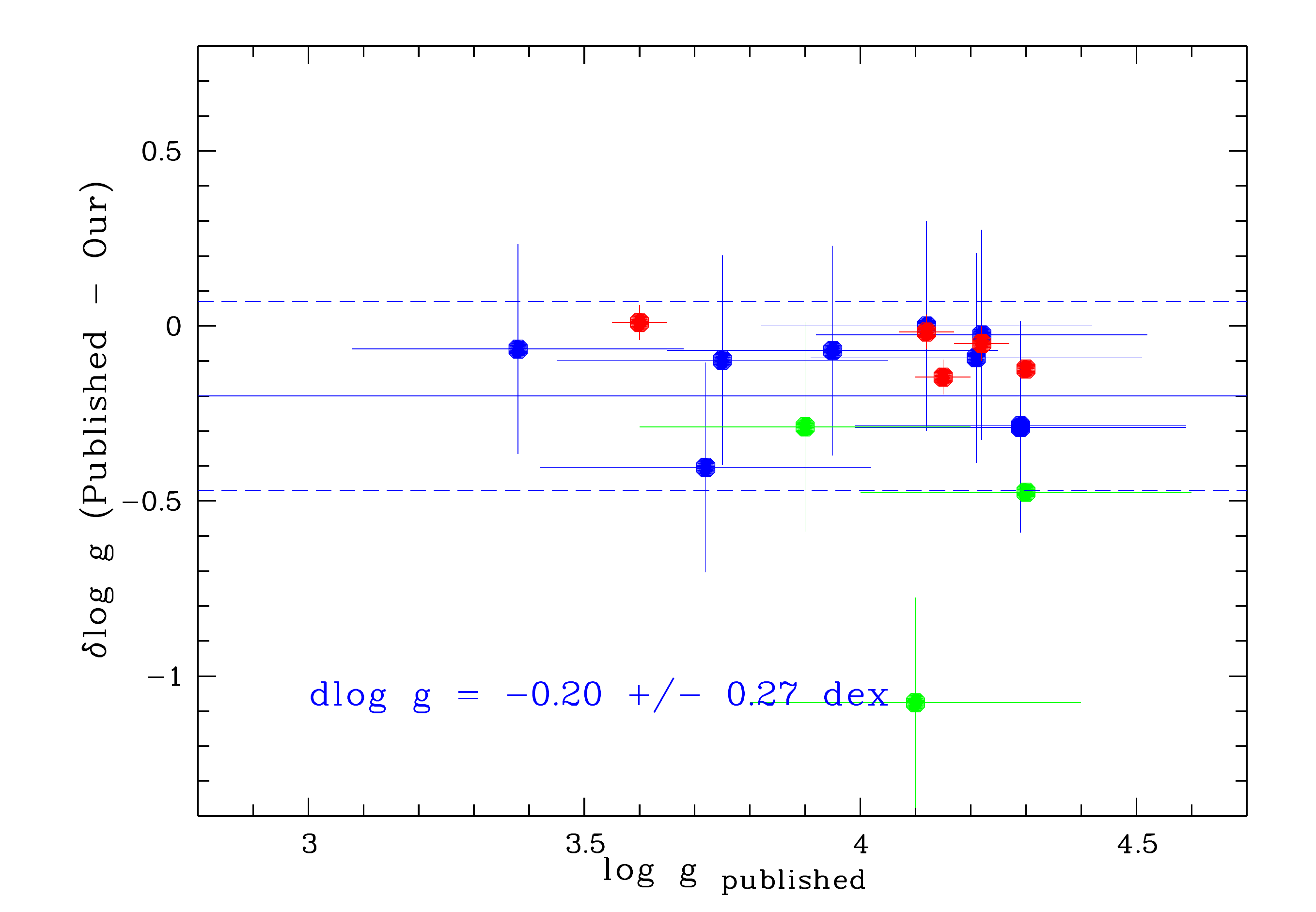}
   \includegraphics[angle=-90, width=\textwidth, clip=]{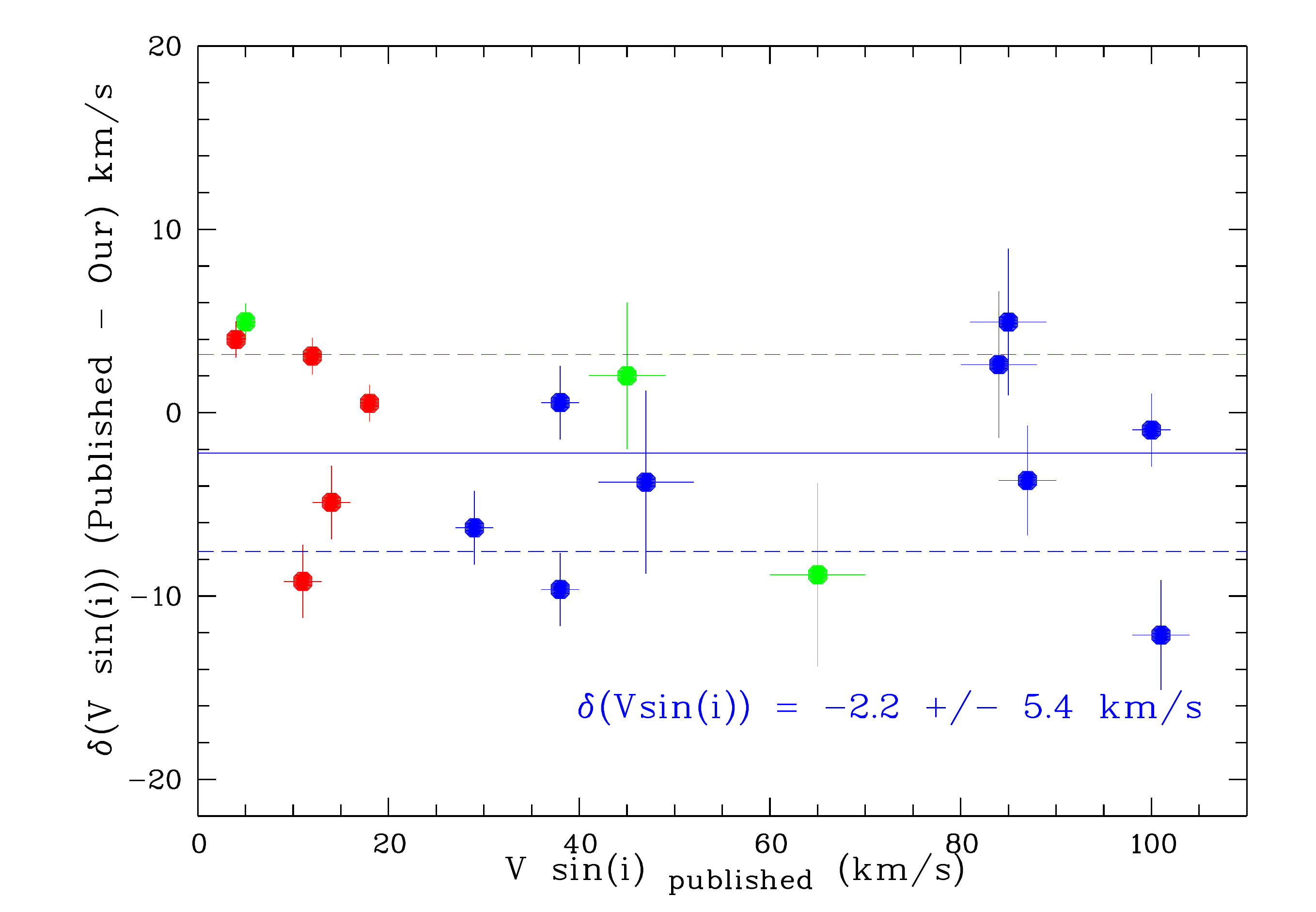}
   \caption{A comparison of the calculated $T_\mathrm{eff}$, $\log g$ and $v \sin i$ with previously published results
   for the sample of early and late B-type stars.}
  \label{fig:Comparison}
 \end{minipage}%
   \end{figure}

Due to the extensive functionality of the \textsc{lmfit} package, the \fbs\ provides flexible
control over model parameters, including upper/lower bounding, parameter fixing and making
parameters connected. As an example, connecting stellar component metallicities
($[Z/H]_1 \equiv [Z/H]_2$) might
be a reasonable approximation for binary stars formed in the same gas cloud.
An example of the use of the \fbs\ software is shown in Figure~\ref{fig:examples} for three
spectra of the binary system FP\,Car from our test sample.
The \fbs\ is a full pixel fitting approach and allows us to approximate the full spectral range
of the given spectrum or one or several spectral intervals as well as
to easily mask bad pixels and/or spectral regions.
The \fbs\ also contains basic functionality to determine orbital parameters from
the stellar rotation curves.

\section{Check-up of the External Accuracy}

To check the external accuracy of our \fbs\ software, we make different tests.
For one of them we fitted with \fbs\ program
18 echelle spectra of early and late B-type stars, that were obtained
with the Fibre-fed Extended Range Optical Spectrograph \citep[FEROS;][]{Kaufer96}
and were modelled and published \citep{HH03,NP12,BL13}.
For this fit we used models from \cite{Coelho14} for the late B-stars
and models from \cite{2003ApJS..146..417L,2007ApJS..169...83L} for the early B-stars.
The comparison of our results for $T_\mathrm{eff}$, $\log g$ and $v \sin i$
with results published earlier are shown in Figure~\ref{fig:Comparison}.
Our found $T_\mathrm{eff}$ is comparable to the previously found with rms of 650~K,
$\log g$ with rms of 0.27~dex and $v \sin i$ with rms of 5.4~km s$^{-1}$
that are very close to cited errors that are shown with gorizontal bars.
There are no any obvious systematic issues are visible in this Figure.

\begin{figure}[t]
\centering{
 \includegraphics[clip=,angle=0,width=0.9\textwidth]{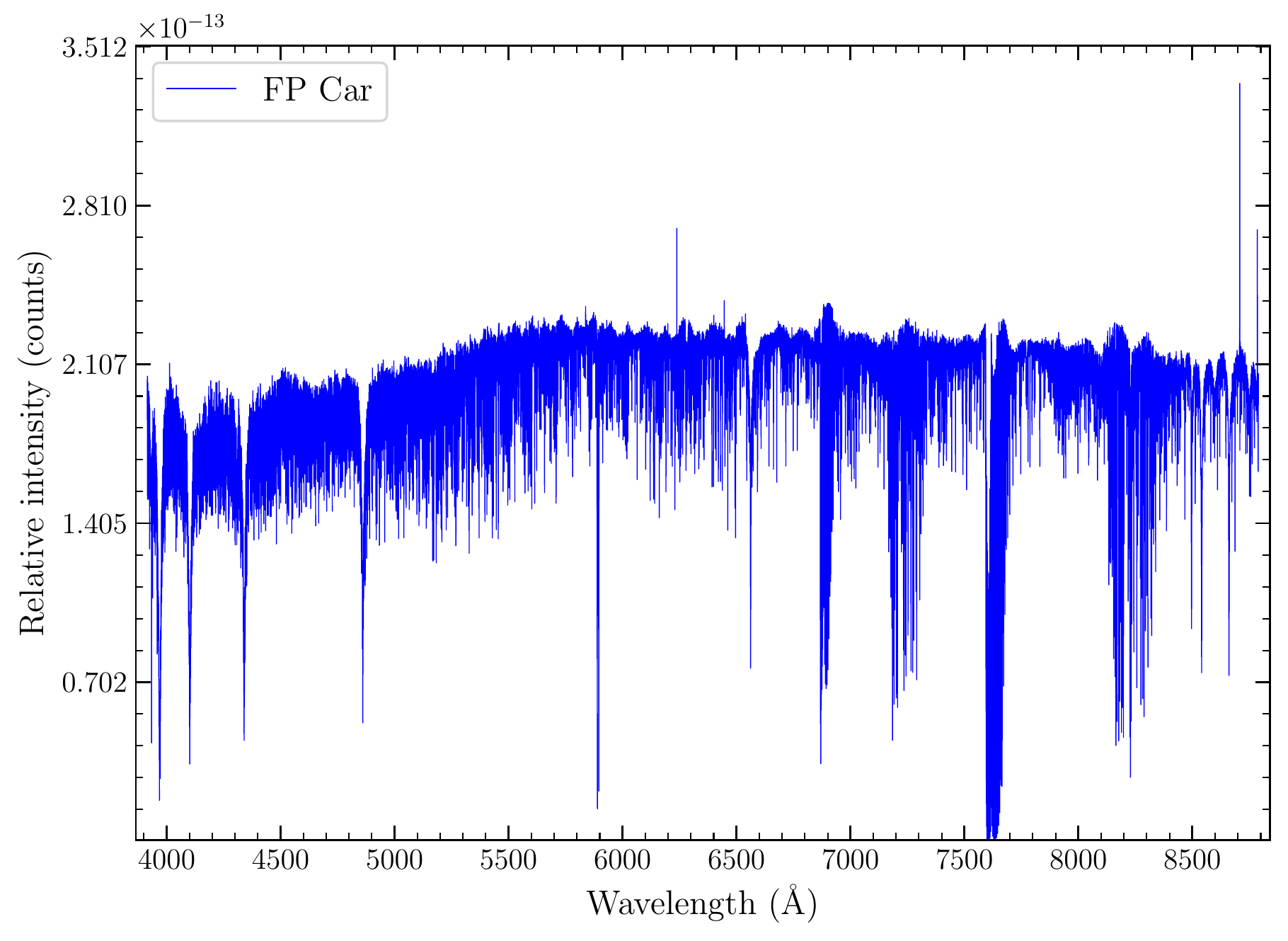}
}
 \caption{An example of a fully processed spectrum of FP\,Car.
The spectrum consists of 70 \'echelle orders from both the blue and red arms
merged together and corrected for the sensitivity curve.
 \label{fig:FP_Car_spec}}
\end{figure}
\begin{figure*}
\centering{
 \includegraphics[clip=,angle=0,width=0.9\textwidth]{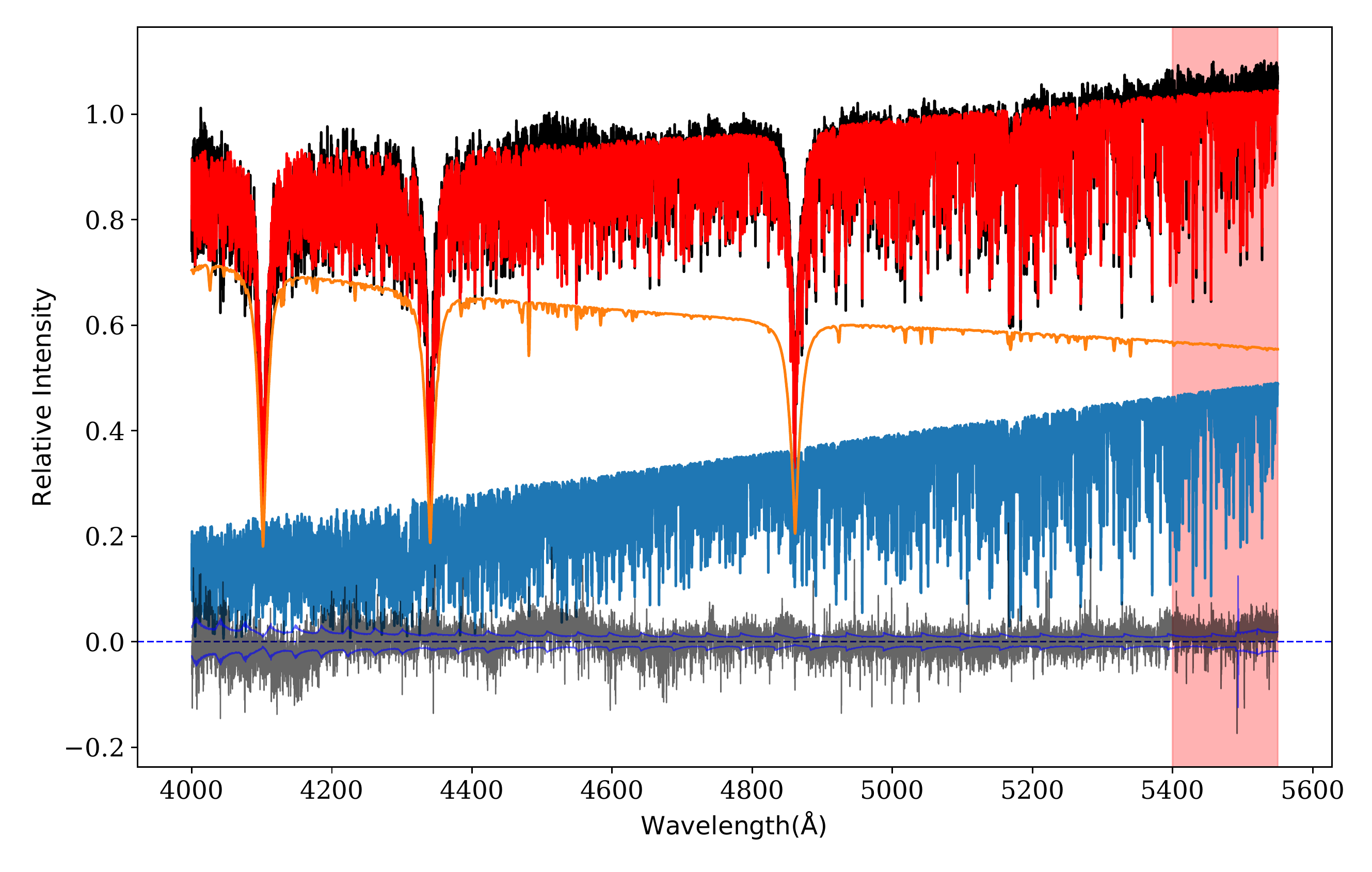}
}
 \caption{The results of the analysis of one spectrum of FP\,Car obtained with HRS. 
     The panel shows the result of the fit in the spectral region 4000-5300~\AA.
     Designations are the same as in Figure~\ref{fig:examples}.
 \label{fig:FP_Car_spec_fit}}
\end{figure*}
\begin{figure}[t]
  \centering
   \includegraphics[angle=0, width=0.9\textwidth, clip=]{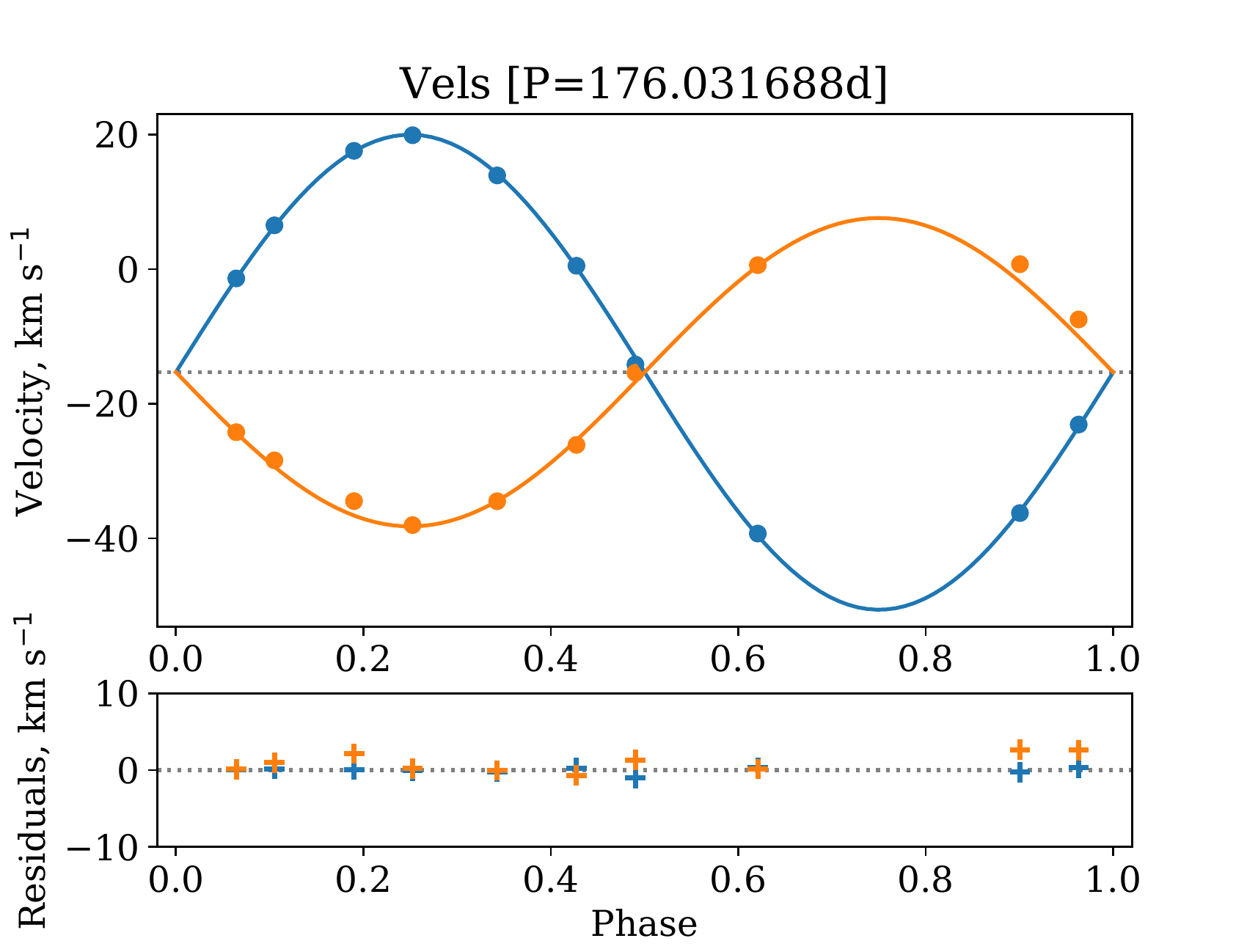}
   \caption{The calculated radial velocity curves for the FP\,Car 
       binary system from our test sample.}
  \label{fig:FP_Car}
   \end{figure}
\begin{figure}[t]
\centering
 \includegraphics[clip=,angle=0,width=0.7\textwidth]{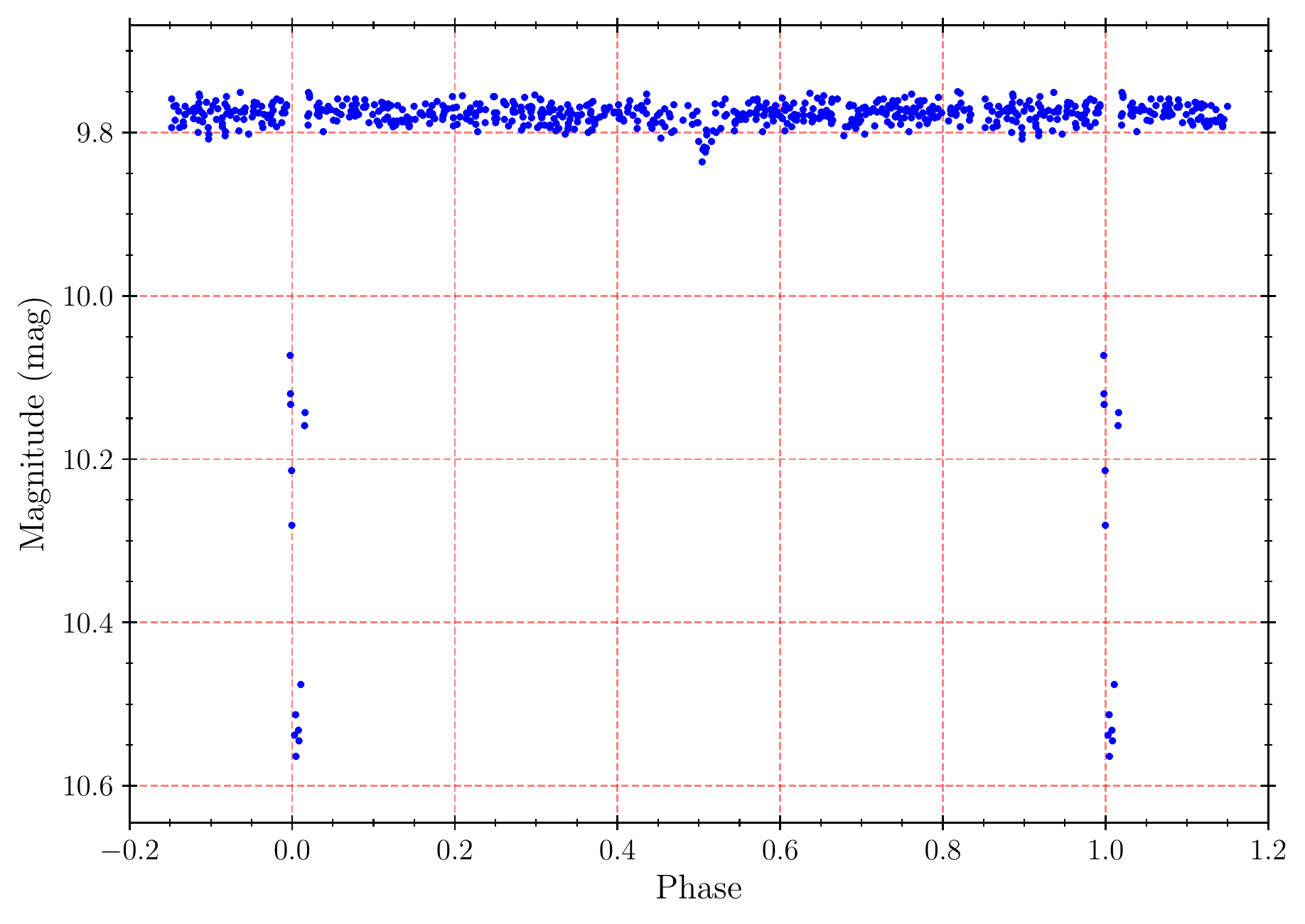}
 \caption{Photometric data from the ASAS survey converted to the period P=176.027~days. 
     There are only a few points that indicate the shape of the primary and secondary minima.
 \label{fig:FP_Car_phot}}
\end{figure}

\begin{table*}
   \centering
   \caption{Best-fit orbital elements.}
   \begin{tabular}{llc}
      \hline\hline
		Parameter                                         &       Value          & \%    \\ \hline
Epoch at radial velocity maximum $T_0$ (d)                & $2455094.47\pm0.15$  & 0.00  \\
Orbital period $P$ (d)                                    & 176.032$\pm$0.010    & 0.00  \\
Eccentricity $e$                                          & 0 (fixed)            & 0.00  \\
Radial velocity semi-amplitude $K1$ (km s$^{-1}$)         & 22.92$\pm$0.73       & 3.20  \\
Radial velocity semi-amplitude $K2$ (km s$^{-1}$)         & 35.30$\pm$0.26       & 0.72  \\
Systemic heliocentric velocity $\gamma$ (km s$^{-1}$)     &$-$15.31$\pm$0.15     & 0.96  \\
Root-mean-square residuals of Keplerian fit (km s$^{-1}$) & 0.976                & --    \\
      \hline
   \end{tabular}
   \label{tab:FP_Car_orb}
\end{table*}

{
\section{First Results: the FP\,Car system}

As the first result we would like to present here some our results on the study of the FP\,Car binary
system that belongs to our test sample (see Table~\ref{Tab:Sample}).
FP\,Car (HD96214) as a variable star with period of $\sim$176~days was discovered 
by \citet{1926BHarO.837....1C}. The period of this system was measured more accurate 
by \citet{2004IBVS.5542....1D} using data from ASAS survey \citep{1997AcA....47..467P}. 
The spectral type was estimated by \citet{1975mcts.book.....H} as B5/7(V). 
Approximate masses $M_1 = 15.61 M_\odot$ and $M_2 = 7.49 M_\odot$ for both components
were calculated by \citet{1980AcA....30..501B} with use their iterative
method for computation of geometric and physical parameters for components of eclipsing binary stars.

Our spectral observations of FP\,Car were made during 2017--2019 with HRS
at SALT (see Section~\ref{sect:Obs}). 
Ten spectra were obtained in total covering all phases of the binary's orbit.
After the standard HRS reduction, additionally, each HRS spectrum of FP\,Car was corrected 
for bad columns and pixels and was also corrected for the spectral sensitivity curve 
obtained closest to the date of the observation. 
Spectrophotometric standards for HRS were observed once a week as a part of the HRS Calibration Plan.
Figure~\ref{fig:FP_Car_spec} presents one of a fully processed spectrum of FP\,Car,
which was used in further analysis.
The spectrum consists of 70 \'echelle orders from both the blue and the red arms of HRS
merged together and corrected for sensitivity.
Unfortunately, SALT is a telescope where the unfilled entrance pupil of the telescope
moves during the observation and for that reason the absolute flux calibration is not
feasible with SALT. At the same time, since all optical elements are always the same, 
the relative flux calibration can be used for SALT data.

All HRS observations of the FP\,Car system were used simultaneously for the calculation 
of radial velocity curves using \fbs\ package. 
The determination of orbital parameters from the stellar rotation curves was also done with \fbs\ package
as it is shown in Figure~\ref{fig:FP_Car} and presented in Table~\ref{tab:FP_Car_orb}.
The found period is P=176.032$\pm$0.010~days that is in agreement within uncertainties with 
photometric period presented in Table~\ref{Tab:Sample}.
Our spectral data show that system has circular orbit (e=0) and this parameter was fixed for the last
iteration. Our found amplitudes of velocities have small errors 0.7\% for the component B (blue) 
and 3.2\% for the component A (orange) that is totally in agreement 
with fit shown in Figure~\ref{fig:examples},
where spectrum of the component B shows many narrow lines and spectrum 
of the component A shows only wide Balmer and helium lines with $v \sin i \sim 100$~km~s$^{-1}$.
Finally, we can calculate masses of both components for the FP\,Car system as
$M_1 = (2.19\pm0.06) \sin^{-3}(i)\,\, M_\odot$ and $M_2 = (1.42\pm0.06) \sin^{-3}(i)\,\, M_\odot$,
where $i$ is the orbital inclination angle that can only be determined from the modeling of
photometric data.
Unfortunately, there is no good photometric data for FP\,Car among all existing public surveys.
The best available data are from the ASAS survey \citep{1997AcA....47..467P} as shown 
in Figure~\ref{fig:FP_Car_phot}. However, even these data 
have too few points outlining the positions and shapes of the narrow primary 
and secondary minima and it is impossible to use these data for any modeling.
For that reason we are actively accumulating photometric data for FP\,Car and
other stars from our test sample using the telescope network LCO \citep{2013PASP..125.1031B}.

\section{Conclusions}
\label{sect:conclusion}

We present our new project on study of the long-period massive eclipsing binaries,
where components are not synchronized and for this reason not changed the evolution scenario
of each other.
Small sample of eleven binary systems was described here that was formed for 
the pilot spectroscopy with HRS/SALT. The software package
\fbs\ (Fitting Binary Stars) was developed by us for the analysis of spectral data.
We describe this package and show its external accuracy in determination stellar parameters.
As the first result we present the radial velocity curves and the best-fit orbital elements
for both components of the FP\,Car binary system from our test sample.
}

\begin{acknowledgements}
All spectral observations reported in this paper were obtained with
the Southern African Large Telescope (SALT)
under programs 2016-1-MLT-002, 2017-1-MLT-001 and 2019-1-SCI-004 (PI: Alexei Kniazev).
AK acknowledges support from the National Research Foundation of South Africa.
OM acknowledges support by the Russian Foundation for Basic Researches grant 20-52-53009.
IK acknowledges support by the Russian Science Foundation grant 17-72-20119.
LB acknowledges support by the Russian Science Foundation grants  18-02-00890 and 19-02-00611.
\end{acknowledgements}

\bibliographystyle{raa}
\bibliography{RAA-2020-0036}

\begin{thebibliography}{48}
\providecommand\natexlab[1]{#1}
\providecommand\JournalTitle[1]{#1}

\bibitem[{Andersen}(1991)]{1991A&ARv...3...91A}
{Andersen}, J. 1991, \aapr, 3, 91

\bibitem[{Avvakumova} \& {Malkov}(2014)]{2014MNRAS.444.1982A}
{Avvakumova}, E.~A., \& {Malkov}, O.~Y. 2014, \mnras, 444, 1982

\bibitem[{Avvakumova} {et~al.}(2013)]{2013AN....334..860A}
{Avvakumova}, E.~A., {Malkov}, O.~Y., \& {Kniazev}, A.~Y. 2013, Astronomische
  Nachrichten, 334, 860

\bibitem[{Bailey} \& {Landstreet}(2013)]{BL13}
{Bailey}, J.~D., \& {Landstreet}, J.~D. 2013, \aap, 551, A30

\bibitem[{Barnes} {et~al.}(2008)]{Ba08}
{Barnes}, S.~I., {Cottrell}, P.~L., {Albrow}, M.~D., {et~al.} 2008, Society of
  Photo-Optical Instrumentation Engineers (SPIE) Conference Series, Vol. 7014,
  {The optical design of the Southern African Large Telescope high resolution
  spectrograph: SALT HRS}, Society of Photo-Optical Instrumentation Engineers
  (SPIE) Conference Series, Vol. 7014, Ground-based and Airborne
  Instrumentation for Astronomy II. Edited by McLean, Ian S.; Casali, Mark M.
  Proceedings of the SPIE, Volume 7014, article id. 70140K, 12 pp. (2008).,
  70140K

\bibitem[{Bramall} {et~al.}(2010)]{Br10}
{Bramall}, D.~G., {Sharples}, R., {Tyas}, L., {et~al.} 2010, Society of
  Photo-Optical Instrumentation Engineers (SPIE) Conference Series, Vol. 7735,
  {The SALT HRS spectrograph: final design, instrument capabilities, and
  operational modes}, Society of Photo-Optical Instrumentation Engineers (SPIE)
  Conference Series, Vol. 7735, Proceedings of the SPIE, Volume 7735, id.
  77354F (2010)., 77354F

\bibitem[{Bramall} {et~al.}(2012)]{Br12}
{Bramall}, D.~G., {Schmoll}, J., {Tyas}, L.~M.~G., {et~al.} 2012, Society of
  Photo-Optical Instrumentation Engineers (SPIE) Conference Series, Vol. 8446,
  {The SALT HRS spectrograph: instrument integration and laboratory test
  results}, Society of Photo-Optical Instrumentation Engineers (SPIE)
  Conference Series, Vol. 8446, Ground-based and Airborne Instrumentation for
  Astronomy IV. Proceedings of the SPIE, Volume 8446, article id. 84460A, 9 pp.
  (2012)., 84460A

\bibitem[{Brancewicz} \& {Dworak}(1980)]{1980AcA....30..501B}
{Brancewicz}, H.~K., \& {Dworak}, T.~Z. 1980, \actaa, 30, 501

\bibitem[{Brown} {et~al.}(2013)]{2013PASP..125.1031B}
{Brown}, T.~M., {Baliber}, N., {Bianco}, F.~B., {et~al.} 2013, \pasp, 125, 1031

\bibitem[{Buckley} {et~al.}(2006)]{Buck06}
{Buckley}, D. A.~H., {Swart}, G.~P., \& {Meiring}, J.~G. 2006, Society of
  Photo-Optical Instrumentation Engineers (SPIE) Conference Series, Vol. 6267,
  {Completion and commissioning of the Southern African Large Telescope},
  Society of Photo-Optical Instrumentation Engineers (SPIE) Conference Series,
  Vol. 6267, Ground-based and Airborne Telescopes. Edited by Stepp, Larry M..
  Proceedings of the SPIE, Volume 6267, id. 62670Z (2006)., 62670Z

\bibitem[{Cannon}(1926)]{1926BHarO.837....1C}
{Cannon}, A.~J. 1926, Harvard College Observatory Bulletin, 837, 1

\bibitem[{Coelho}(2014)]{Coelho14}
{Coelho}, P.~R.~T. 2014, \mnras, 440, 1027

\bibitem[{Crause} {et~al.}(2014)]{Cr14}
{Crause}, L.~A., {Sharples}, R.~M., {Bramall}, D.~G., {et~al.} 2014, Society of
  Photo-Optical Instrumentation Engineers (SPIE) Conference Series, Vol. 9147,
  {Performance of the Southern African Large Telescope (SALT) High Resolution
  Spectrograph (HRS)}, Society of Photo-Optical Instrumentation Engineers
  (SPIE) Conference Series, Vol. 9147, Proceedings of the SPIE, Volume 9147,
  id. 91476T 14 pp. (2014)., 91476T

\bibitem[{Crawford} {et~al.}(2010)]{Cra2010}
{Crawford}, S.~M., {Still}, M., {Schellart}, P., {et~al.} 2010, Society of
  Photo-Optical Instrumentation Engineers (SPIE) Conference Series, Vol. 7737,
  {PySALT: the SALT science pipeline}, Society of Photo-Optical Instrumentation
  Engineers (SPIE) Conference Series, Vol. 7737, Proceedings of the SPIE,
  Volume 7737, id. 773725 (2010)., 773725

\bibitem[{Delfosse} {et~al.}(2000)]{2000A&A...364..217D}
{Delfosse}, X., {Forveille}, T., {S{\'e}gransan}, D., {et~al.} 2000, \aap, 364,
  217

\bibitem[{Docobo} {et~al.}(2016)]{2016MNRAS.459.1580D}
{Docobo}, J.~A., {Tamazian}, V.~S., {Malkov}, O.~Y., {Campo}, P.~P., \&
  {Chulkov}, D.~A. 2016, \mnras, 459, 1580

\bibitem[{Dvorak}(2004)]{2004IBVS.5542....1D}
{Dvorak}, S.~W. 2004, Information Bulletin on Variable Stars, 5542, 1

\bibitem[{Fernandes} {et~al.}(1998)]{1998A&A...338..455F}
{Fernandes}, J., {Lebreton}, Y., {Baglin}, A., \& {Morel}, P. 1998, \aap, 338,
  455

\bibitem[{Gorda} \& {Svechnikov}(1998)]{1998ARep...42..793G}
{Gorda}, S.~Y., \& {Svechnikov}, M.~A. 1998, Astronomy Reports, 42, 793

\bibitem[{Gray}(1992)]{Gray92}
{Gray}, D.~F. 1992, {The observation and analysis of stellar photospheres.},
  Vol.~20

\bibitem[{Hempel} \& {Holweger}(2003)]{HH03}
{Hempel}, M., \& {Holweger}, H. 2003, \aap, 408, 1065

\bibitem[{Henry}(2004)]{2004ASPC..318..159H}
{Henry}, T.~J. 2004, in Astronomical Society of the Pacific Conference Series,
  Vol. 318, Spectroscopically and Spatially Resolving the Components of the
  Close Binary Stars, ed. R.~W. {Hilditch}, H.~{Hensberge}, \& K.~{Pavlovski},
  159

\bibitem[{Henry} {et~al.}(1999)]{1999ApJ...512..864H}
{Henry}, T.~J., {Franz}, O.~G., {Wasserman}, L.~H., {et~al.} 1999, \apj, 512,
  864

\bibitem[{Houk} \& {Cowley}(1975)]{1975mcts.book.....H}
{Houk}, N., \& {Cowley}, A.~P. 1975, {University of Michigan Catalogue of
  two-dimensional spectral types for the HD stars. Volume I. Declinations -90.
  to -53.}

\bibitem[{Husser} {et~al.}(2013)]{phoenix}
{Husser}, T.~O., {Wende-von Berg}, S., {Dreizler}, S., {et~al.} 2013, \aap,
  553, A6

\bibitem[{Kaufer} {et~al.}(1996)]{Kaufer96}
{Kaufer}, A., {Stahl}, O., {Wolf}, B., {et~al.} 1996, \aap, 305, 887

\bibitem[{Khaliullin} \& {Khaliullina}(2007)]{2007MNRAS.382..356K}
{Khaliullin}, K.~F., \& {Khaliullina}, A.~I. 2007, \mnras, 382, 356

\bibitem[{Khaliullin} \& {Khaliullina}(2010)]{2010MNRAS.401..257K}
{Khaliullin}, K.~F., \& {Khaliullina}, A.~I. 2010, \mnras, 401, 257

\bibitem[{Kniazev} {et~al.}(2016)]{KGB16}
{Kniazev}, A.~Y., {Gvaramadze}, V.~V., \& {Berdnikov}, L.~N. 2016, \mnras, 459,
  3068

\bibitem[{Kniazev} {et~al.}(2019)]{KUKB19}
{Kniazev}, A.~Y., {Usenko}, I.~A., {Kovtyukh}, V.~V., \& {Berdnikov}, L.~N.
  2019, Astrophysical Bulletin, 74, 208

\bibitem[{Kovaleva}(2001)]{2001ARep...45..972K}
{Kovaleva}, D.~A. 2001, Astronomy Reports, 45, 972

\bibitem[{Lanz} \& {Hubeny}(2003)]{2003ApJS..146..417L}
{Lanz}, T., \& {Hubeny}, I. 2003, \apjs, 146, 417

\bibitem[{Lanz} \& {Hubeny}(2007)]{2007ApJS..169...83L}
{Lanz}, T., \& {Hubeny}, I. 2007, \apjs, 169, 83

\bibitem[{Malkov}(2003)]{2003A&A...402.1055M}
{Malkov}, O.~Y. 2003, \aap, 402, 1055

\bibitem[{Malkov}(2007)]{2007MNRAS.382.1073M}
{Malkov}, O.~Y. 2007, \mnras, 382, 1073

\bibitem[{Malkov} {et~al.}(2007)]{2007A&A...465..549M}
{Malkov}, O.~Y., {Oblak}, E., {Avvakumova}, E.~A., \& {Torra}, J. 2007, \aap,
  465, 549

\bibitem[{Malkov} {et~al.}(1997)]{1997A&A...320...79M}
{Malkov}, O.~Y., {Piskunov}, A.~E., \& {Shpil'Kina}, D.~A. 1997, \aap, 320, 79

\bibitem[{Malkov} {et~al.}(2012)]{2012A&A...546A..69M}
{Malkov}, O.~Y., {Tamazian}, V.~S., {Docobo}, J.~A., \& {Chulkov}, D.~A. 2012,
  \aap, 546, A69

\bibitem[{Newville} {et~al.}(2016)]{lmfit}
{Newville}, M., {Stensitzki}, T., {Allen}, D.~B., {et~al.} 2016, {Lmfit:
  Non-Linear Least-Square Minimization and Curve-Fitting for Python}

\bibitem[{Nieva} \& {Przybilla}(2012)]{NP12}
{Nieva}, M.~F., \& {Przybilla}, N. 2012, \aap, 539, A143

\bibitem[{O'Donoghue} {et~al.}(2006)]{Dono06}
{O'Donoghue}, D., {Buckley}, D.~A.~H., {Balona}, L.~A., {et~al.} 2006, \mnras,
  372, 151

\bibitem[{Pojmanski}(1997)]{1997AcA....47..467P}
{Pojmanski}, G. 1997, \actaa, 47, 467

\bibitem[{Popper}(1980)]{1980ARA&A..18..115P}
{Popper}, D.~M. 1980, \araa, 18, 115

\bibitem[{Tassoul}(1987)]{1987ApJ...322..856T}
{Tassoul}, J.-L. 1987, \apj, 322, 856

\bibitem[{Tassoul}(1988)]{1988ApJ...324L..71T}
{Tassoul}, J.-L. 1988, \apjl, 324, L71

\bibitem[{Torres} {et~al.}(2010)]{2010A&ARv..18...67T}
{Torres}, G., {Andersen}, J., \& {Gim{\'e}nez}, A. 2010, \aapr, 18, 67

\bibitem[{Zahn}(1975)]{1975A&A....41..329Z}
{Zahn}, J.~P. 1975, \aap, 41, 329

\bibitem[{Zahn}(1977)]{1977A&A....57..383Z}
{Zahn}, J.~P. 1977, \aap, 500, 121

\end{thebibliography}

\label{lastpage}

\end{document}